\documentclass[aps,twocolumn, secnumarabic,nobalancelastpage,amsmath,amssymb,nofootinbib]{revtex4}

\usepackage{lgrind} 
\usepackage{chapterbib}
\usepackage{color}
\usepackage{graphics} 
\usepackage{graphicx} 
\usepackage{subfigure}
\usepackage{longtable} 
\usepackage{bm} 
\usepackage[colorlinks=true]{hyperref}
\usepackage{epstopdf}
\newcommand{\bra}[1]{\ensuremath{\left\langle{#1}\right\vert}}
\newcommand{\ket}[1]{\ensuremath{\left\vert{#1}\right\rangle}}

\newcommand{\expect}[1]{\ensuremath{\left\langle{#1}\right\rangle}}
\newcommand{\ident}{{\mathbf{1}}}

\newcommand{\beq}{\begin{equation}}
\newcommand{\eeq}{\end{equation}}
\newcommand{\bqa}{\begin{eqnarray}}
\newcommand{\eqa}{\end{eqnarray}}
\newcommand{\nn}{\nonumber}
\newcommand{\eg}{\emph{e.g.},~}
\newcommand{\ie}{\emph{i.e.},~}
\newcommand{\dg}{^\dagger}

\newcommand{\brho}{\bar{\rho}}
\newcommand{\bpsi}{\bar{\psi}}
\newcommand{\Dt}{\Delta t}
\newcommand{\Dw}{{\Delta W}}

\newcommand{\BQIC}{Berkeley Center for Quantum Information and Computation, Berkeley, California 94720 USA}
\newcommand{\DeptPhys}{Department of Physics, University of California, Berkeley, California 94720 USA}
\newcommand{\DeptChem}{Department of Chemistry, University of California, Berkeley, California 94720 USA}
\newcommand{\SNL}{Digital \& Quantum Information Systems, Sandia National Laboratories, Livermore, California 94550 USA}

\newcommand{\erf}[1]{Eqn.~(\ref{#1})}

\begin{document}

\title{Deterministic generation of remote entanglement with active quantum feedback}

\author {Leigh Martin$^{1,2}$}
\email {Leigh@Berkeley.edu}
\author{Felix Motzoi$^{1,2,3}$}
\thanks{Current address: Theoretical Physics, Saarland University, 66123 Saarbr{\"u}cken, Germany}
\author{Hanhan Li$^{1,2}$}
\author{Mohan Sarovar$^{4}$}
\author{K. Birgitta Whaley$^{1,3}$}

\address{$^1$\BQIC}
\address{$^2$\DeptPhys}
\address{$^3$\DeptChem}
\address{$^4$\SNL}

\date{\today}

\begin{abstract}
We consider the task of deterministically entangling two remote qubits using joint measurement and feedback, but no directly entangling Hamiltonian. In order to formulate the most effective experimentally feasible protocol, we introduce the notion of average sense locally optimal (ASLO) feedback protocols, which do not require real-time quantum state estimation, a difficult component of real-time quantum feedback control. We use this notion of optimality to construct two protocols which can deterministically create maximal entanglement: a semiclassical feedback protocol for low efficiency measurements and a quantum feedback protocol for high efficiency measurements. The latter reduces to direct feedback in the continuous-time limit, whose dynamics can be modeled by a Wiseman-Milburn feedback master equation which yields an analytic solution in the limit of unit measurement efficiency. Our formalism can smoothly interpolate between continuous-time and discrete-time descriptions of feedback dynamics, and we exploit this feature to then derive a superior hybrid protocol for arbitrary non-unit measurement efficiency that switches between quantum and semiclassical protocols. Finally, we show using simulations incorporating experimental imperfections that deterministic entanglement of remote superconducting qubits may be achieved with current technology using the continuous-time feedback protocol alone.

\end{abstract}

\maketitle

\section{Introduction}
\label{sec:introduction}

Engineering of quantum devices requires optimization of two essentially contradictory requirements. On one hand, quantum properties such as superposition and entanglement upon which these devices rely are fragile, and require careful isolation from external degrees of freedom. On the other hand, control and measurement of a system requires coupling to an external device, which often runs contrary to the necessity for decoupling from the environment. 
To balance this trade-off, many quantum systems relevant for quantum computing and sensing lack readout capabilities that are effectively instantaneous and projective.
Instead, measurement occurs over a finite, resolvable time scale in such systems. Recent research has taken advantage of this by utilizing the fact that continuous weak measurement enables direct observation of the continuous time evolution of a quantum system (quantum trajectories) \cite{Car-1993a, Hoo.Lyn.etal-2000, Murch:2013ur, Roch:2014ey}, and also permits operations on the system \emph{during} the measurement process, including feedback and feed-forward control \cite{Wiseman:2009vw, Rei.Smi.etal-2004, Sayrin:2011jx, Vijay:2012ua}.

Real-time quantum feedback control is expected to be broadly applicable to many problems in quantum information science. Some quantum information applications that have been proposed to date include, rapid purification of qubits or qubit registers \cite{Jacobs:2003hc,Wiseman:2006ti, Combes:2006hx, Wiseman:2008bc, Combes:2011wt, Li:2013vd}, quantum error correction \cite{Ahn.Doh.etal-2002, Sar.Ahn.etal-2004}, transmission of quantum information through noisy channels \cite{Branczyk:2007dl}, adaptive measurement for quantum state discrimination \cite{Wis-1995, Coo.Mar.etal-2007, Sar.Wha-2007}, and several forms of quantum state preparation and stabilization \cite{Mab.Zol-1996, Wan.Wis-2001,Ste.Jac.etal-2004,Sto.Han.etal-2004,Han.Sto.etal-2005}.

Experimentally, quantum feedback has 
been mostly demonstrated to date with atomic, molecular, or optical systems, \eg \cite{Rei.Smi.etal-2004, Sayrin:2011jx}. However, the 
advent of quantum-limited microwave amplifiers has recently enabled experimental realization of quantum feedback in superconducting circuits, where it has been used to stabilize the Rabi oscillations of a single qubit \cite{Vijay:2012ua} and to deterministically create entanglement of two qubits within a single cavity \cite{Riste:2013um} using 
a discrete feedback loop.

Several works have also suggested using quantum feedback to enhance generation of remote entanglement \cite{Wan.Wis.etal-2004, Sar.Goa.etal-2005, Kerckhoff:2009ch, Liu:2010kd}. These 
papers have considered the case in which the controller has access to a joint measurement on a pair of qubits that are too far apart to engineer a direct or photon-mediated interaction, an important scenario for quantum networks or large-scale quantum computers \cite{Benjamin:2013}. The remote aspect requires that 
the quantum feedback operations be restricted to local unitaries, which cannot on their own generate entanglement \cite{Nielsen1998}. Conversely, a joint measurement alone cannot deterministically project a separable system into a fixed entangled state, but access to this measurement and local unitary feedback can~\cite{lloyd2001engineering}.

In this work, we build upon 
this literature in our goal to achieve optimal protocols for remote entanglement, focusing in particular on joint measurements that have been  implemented in superconducting qubits \cite{Roch:2014ey}. We motivate our approach by an examination of optimality for the state update in a single, discrete time step and define an \emph{average sense locally optimal}  (ASLO) strategy that makes the state update over this discrete time step using the most recent measurement outcome and knowledge of the average state. We show that this discrete time step protocol 
reduces to a direct feedback protocol in the continuous-time limit, whose dynamics can be modeled by a Wiseman-Milburn feedback master equation with an analytic solution that yields a simple, closed-form expression for the locally optimal quantum feedback in this limit.  This
analytic solution shows that for unit measurement efficiency, the fidelity with respect to the target entangled state asymptotes exponentially to $1$.  The discrete time step optimality study also motivates a semiclassical protocol that 
is more effective for low measurement efficiencies,
provided that one takes longer time steps between applications of feedback.
This feedback strategy is capable of fully entangling the qubits even for non-unit efficiency, but creates entanglement more slowly than the continuous case.
Both semiclassical and quantum protocols asymptote to unit entanglement fidelity under unit measurement efficiency. 
In the case of non-ideal measurements, the preferred form of the optimal feedback protocol is found to depend on the measurement efficiency $\eta$, with the semiclassical/quantum protocol being preferred for low/high $\eta$ values.  
We then show that for arbitrary non-unit efficiency, continuously tuning the measurement time leads to a protocol that can surpass this ASLO strategy and that switches from using the quantum protocol at short times to the semiclassical protocol at larger times.

The remainder of the paper is organized as follows. In section \ref{sec:Measurement}, we describe the measurement that conditions our feedback
using both stochastic master equation and POVM descriptions. 
 Section \ref{sec:OptFeedback} derives the optimal feedback unitary 
for a single discrete time step as a function of the prior state of the qubits. 
We first consider the case of inefficient measurement in section \ref{sec:ThresholdFeedback}, where we neglect coherence terms of the density matrix that are dephased by the measurement to obtain a semiclassical strategy. 
In section \ref{sec:ContinuousFeedback}, we use the full density matrix for the system to derive a quantum strategy that is valid in the continuous measurement limit. 
In section \ref{sec:Changedt} we combine the semiclassical and quantum protocols to develop a superior hybrid protocol that allows unit fidelity to be reached for arbitrary non-unit measurement efficiencies. Section \ref{sec:Experiment} demonstrates the experimental feasibility of the continuous-time ASLO protocol with numerical simulations of remote entanglement generation in a realizable system of transmon qubits in spatially separated cavities. 
Section \ref{sec:Conclusion} summarizes and provides an outlook for future work.

\section{Entanglement Via Measurement}
\label{sec:Measurement}

To study feedback in the context of remote entanglement generation, we must first describe the measurement 
on which the feedback will rely. 
As analyzed in \cite{Mot.Sar:2015}, it is possible using a dispersive homodyne readout to implement a \emph{half-parity measurement} in the superconducting circuit architecture. In reference \cite{Roch:2014ey}, authors succeeded in applying this measurement to two superconducting transmon qubits separated by over a meter. 
This measurement is characterized by the operator

\begin{equation}
\hat{X} = \frac{1}{2}(\sigma_{z1}+\sigma_{z2}),
\label{eq:HalfParityMeasurement}
\end{equation}
where $\sigma_{z i}$ is the Pauli operator acting on the $i$th qubit. 
This can be used to probabilistically generate entanglement by first preparing the separable uniform superposition state 
$\ket{\psi_0}=\frac{1}{2}(\ket{00} + \ket{01} + \ket{10} + \ket{11})$ (e.g., by making two local $\sigma_y$ rotations on the ground state $\ket{11}$) and then measuring $\hat{X}$. Since this observable cannot distinguish the states $|01\rangle$ and $|10\rangle$, 
such a measurement (if perfect) will with 50 percent probability project the initially unentangled $|\psi_0\rangle$ 
into the entangled triplet state $\ket{t0} = \frac{1}{\sqrt{2}}(|01\rangle + |10\rangle)$.

\begin{figure}[h]
    \centering
        \includegraphics[width=0.45\textwidth]{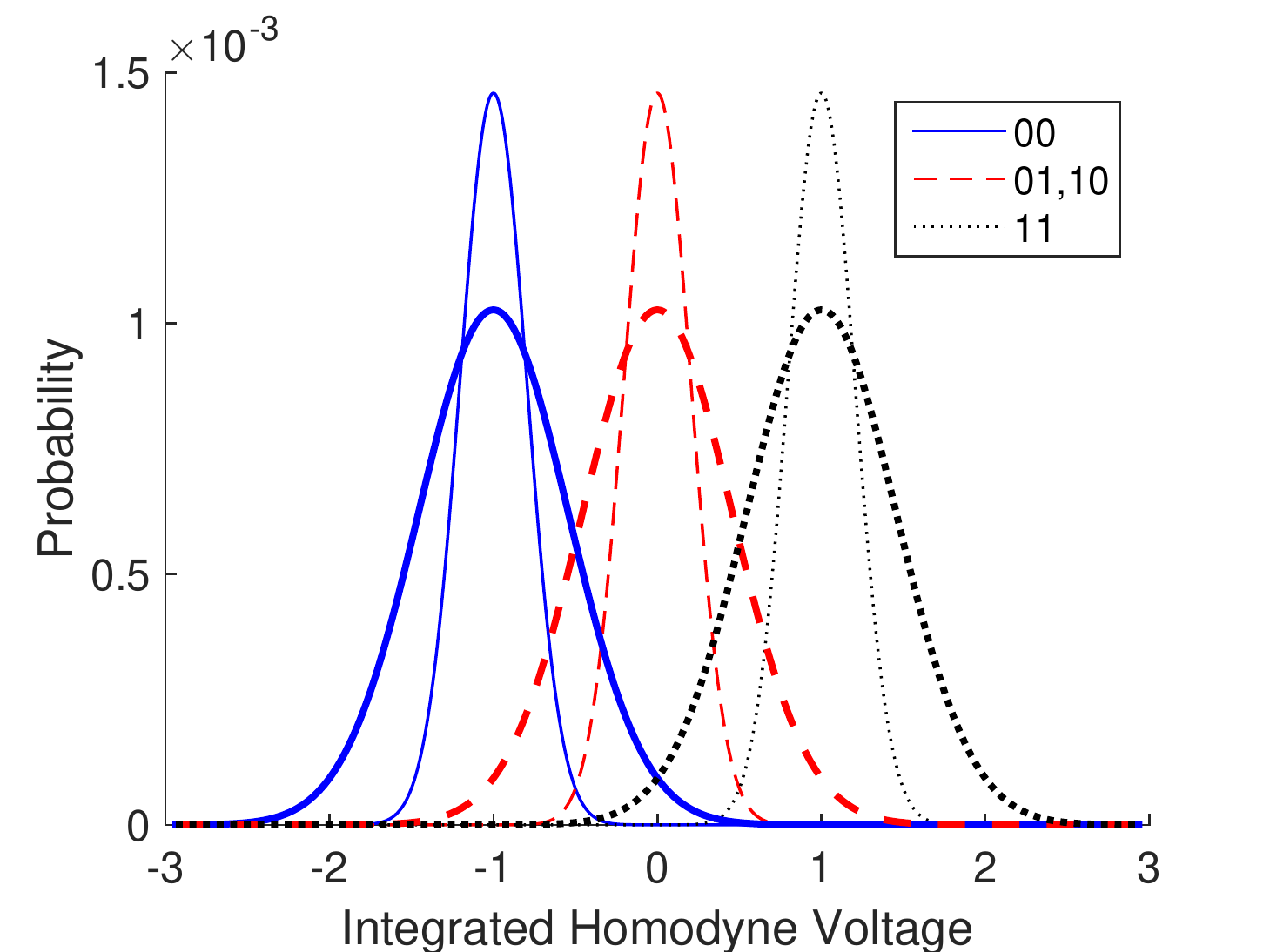}
    \caption{ 
Histograms of the integrated measurement signal for the three eigenstates $\rho_i$ of the measurement, given by $\text{Tr}[\Omega_{\Delta V,\eta} ~\rho_i ~\Omega_{\Delta V, \eta}^\dagger]$.
Plots are for 
      for $\eta=1$ (thin lines) and $\eta = 0.2$ (thick lines) and use the parameters $\Delta t = 3 \mu s$ and $k = 1 ~2\pi\text{MHz}$.}
   \label{fig:Geometry}
\end{figure}

In the superconducting qubit architecture, the implementation of the measurement is such that 
 projection onto one of the eigenstates of $\hat X$ can be said to occur only after a resolvable time period.  During this time interval one can then perform feedback on the system based on information already gained. The measured value of the observable $\hat{X}$ is obtained from a homodyne measurement of the voltage, $V$~\cite{Roch:2014ey}.  During the measurement, the time-dependent homodyne voltage that constitutes the measurement signal is specified by \cite{JacobsSteck2006}
\begin{equation}
dV_t=\langle X \rangle(t) dt + \frac{dW(t)}{\sqrt{8 \eta k}},
\label{eq:HomV}
\end{equation}
where $\expect{\cdot}$ denotes an expectation value under $\rho(t)$, the state of the two qubit system, $k$ is the strength of the measurement, $0\leq\eta \leq 1$ is the measurement efficiency and $dW(t)$ is a Wiener increment satisfying $dW(t)dW(t') = \delta(t-t') dt$ that represents the quantum noise in the homodyne detection~\cite{Wiseman:2009vw}. Note that we have chosen $dV$ to have units of time, so that the average signal $\Delta V = \frac{1}{\Dt}\int dV$ is unitless. Experimentally, the appropriate scale factor can be extracted simply by preparing states with $\langle X\rangle = \{1,0,-1\}$ and measuring $\langle \Delta V \rangle$ for any pair of these states.

The state of the system conditioned on this continuous measurement record is given by the following stochastic master equation~\cite{Wiseman:2009vw}:
\begin{align}
&\rho_{dV}(t+dt) = \rho(t) + \mathcal{D}[M]\rho(t) dt + \mathcal{H}[M] \rho(t) \sqrt{\eta} dW(t); \nonumber \\
&M \equiv \sqrt{2k}X,
\label{eq:RhoMCont}
\end{align}
Here $\mathcal{D}[M] \equiv M \rho M^\dagger -1/2(M^\dagger M \rho + \rho M^\dagger M)$ and $\mathcal{H}[M]\rho \equiv M \rho + \rho M^\dagger - \langle M+M^\dagger \rangle \rho$
(throughout this paper we use units such that $\hbar=1$). The second term in Eq.~(\ref{eq:RhoMCont}) represents the dephasing due to the measurement and the third term represents the information update of the system state derived from the measurement outcome.  Note that equation~\ref{eq:RhoMCont} is expressed in the interaction picture with respect to the free Hamiltonian for the two qubit system, $H_0 = -\frac{\omega_1}{2}\sigma_{z1} - \frac{\omega_2}{2}\sigma_{z2}$.

Eq.~(\ref{eq:RhoMCont}) generates a set of stochastic {\em quantum trajectory} equations that provides a description of the measurement conditioned dynamics in continuous time. In the following we will also be interested in the conditioned dynamics at discrete time intervals.  In order to obtain such a discretized description of the system, we must compute the finite-time generalized measurement, or positive operator valued measure (POVM) \cite{Kra-1983,Wiseman:2009vw} that is associated with the above continuous time weak measurement. Such finite-time POVM descriptions are generally difficult to compute for weak measurements. However in the case of quantum non-demolition (QND) measurements, for which $[H_0, X]=0$, this can be readily derived. As shown in Appendix \ref{app:effect}, the finite time POVM for such weak QND measurements is composed of a set of \emph{effects} $E(\Delta V) = \Omega_{\Delta V}\dg \Omega_{\Delta V}$~\cite{Kra-1983}, with $\Omega_{\Delta V}$ an 
operator corresponding to the measurement of voltage increment $\Delta V$ during a finite time interval $\Delta t$. These effects satisfy the completion property $\int E(\Delta V) d(\Delta V) \equiv \int \Omega_{\Delta V}\dg \Omega_{\Delta V} d(\Delta V) = \ident$, where the integral is performed over the domain of the voltage increment. As shown in Appendix \ref{app:effect}, in the case of perfect measurement efficiency, $\eta=1$, the operator corresponding to the above weak measurement is given by
\begin{align}
\Omega_{\Delta V} = \Bigg( \frac{4 k \Delta t}{\pi} \Bigg)^{1/4} \exp \left[-2 k \Delta t (\Delta V - \hat{X})^2 \right].
\label{eq:MOp_eta1}
\end{align}
The state update corresponding to a discrete time measurement outcome $\Delta V$ is then given by the familiar relation $\rho \rightarrow \Omega_{\Delta V} \rho \Omega^\dagger_{\Delta V}/\text{Tr}[\Omega_{\Delta V} \rho \Omega^\dagger_{\Delta V}]$.
In order to model inefficient measurement, we note that a weak QND measurement with strength $k$ and efficiency $\eta\neq 1$ can be viewed as a sequence of two weak measurements of the same observable, the first with strength $k\eta$ and the second with strength $k(1-\eta)$, where we integrate out the latter in order to model the loss of this portion of the measurement signal. The measurement operator corresponding to the observed fraction $\eta$ of the signal is given by 
\begin{align}
\Omega_{\Delta V, \eta} = \Bigg( \frac{4 \eta k \Delta t}{\pi} \Bigg)^{1/4} \exp \left[-2 \eta k \Delta t (\Delta V - \hat{X})^2 \right].
\label{eq:MOp}
\end{align}

After evolution by \erf{eq:RhoMCont} for a discrete time $\Delta t$, the two-qubit state conditioned on the finite voltage increment $\Delta V$ is then given by 
\begin{align}
&\rho_{\Delta V}(t+\Delta t) = \nonumber \\
& \int_{-\infty}^\infty \Omega_{\Delta V',1-\eta} \frac{\Omega_{\Delta V,\eta} ~\rho(t)~ \Omega_{\Delta V,\eta}^\dagger}{\text{Tr}[ \Omega_{\Delta V,\eta} ~\rho(t)~ \Omega_{\Delta V,\eta}^\dagger]} \Omega_{\Delta V',1-\eta}^\dagger d(\Delta V'). 
\label{eq:RhoM}
\end{align}
Equations~\ref{eq:RhoMCont} and~\ref{eq:RhoM} define the time evolution of $\rho$ under generalized measurement of the observable $X$ for continuous and discrete time increments, respectively. In figure \ref{fig:Geometry}, we 
plot 
histograms of the measurement outcomes for 
$\eta=1$ and $\eta = 0.2$, at a fixed measurement time $\Delta t$. For an inefficient measurement, the variance of the outcomes is larger, which represents uncertainty due to loss or noise. For unit efficiency, there is a residual variance due to vacuum fluctuations of the readout signal, and overlap between the Gaussians due to this effect leads to the measurement being non-projective, or weak. The limit of projective measurement can be recovered by either letting the $k \rightarrow \infty$, i.e., making the measurement infinitely strong, or $\Delta t \rightarrow \infty$, corresponding to an infinitely long measurement time. As noted earlier, putting the system initially into the equal superposition state $\psi_0$ and then measuring $X$ can project the system into a maximally entangled state, but the probability for achieving this by measurement alone cannot exceed 50\%. In the following sections, we will see that feedback can increase this probability to $1$.

\section{Optimal Rotation In a Single Time Step}
\label{sec:OptFeedback}

The only nonlocal resource in our deterministic entanglement scheme is the half-parity measurement $X$ that was introduced in the previous section. Due to the lack of direct qubit-qubit coupling, the remaining control resource, the quantum feedback operations, will be restricted to local rotations that act on each qubit individually. 
Specifically, we define the feedback unitary as
 \begin{equation}
 U_F[\theta_1, \theta_2, \phi_1, \phi_2] = U(\theta_1, \phi_1)\otimes U(\theta_2, \phi_2)
 \label{eq:Unitary}
 \end{equation}
 where 
$$
 U(\theta, \phi) \equiv \hat{I}_2 \cos \theta/2 - i \mathbf{\hat{n}}(\phi) \cdot \mathbf{\hat{\sigma}} \sin \theta/2 \nonumber
$$
is a general single qubit unitary rotation ($\mathbf{\hat{\sigma}}$ is a 3-vector of the Pauli matrices, $\mathbf{\hat{n}}(\phi)$ is a real 3-vector of unit norm and $\hat{I}_2$ is the identity matrix). 
Given the starting point of the equal superposition state $|\psi_0\rangle$, 
any target state lying within the triplet manifold may be obtained by rotations within a fixed $\phi$ plane of the Bloch sphere of each qubit. Hence we set $\mathbf{\hat{n}}(\phi)\cdot \mathbf{\hat{\sigma}} = \sigma_x \cos \phi + \sigma_y \sin \phi $. 
We note that in the presence of dephasing one might wish to remove this restriction and additionally allow $\sigma_z$ rotations, in order to introduce rotations between the target state $\ket{t0}\equiv \frac{1}{\sqrt{2}}(|01\rangle+|10\rangle)$ and the corresponding singlet $\ket{s} = \frac{1}{\sqrt{2}}(|01\rangle - |10\rangle)$.  However, our measurement operator $X$ is not capable of distinguishing these two states, and so does not yield any direct information determining when to apply this operation. Hence the present construction of optimal strategies will assume negligible decoherence, although we shall retain the density matrix element $\rho_{ss}$ throughout the analysis in order to study the impact of dephasing (see section \ref{sec:Experiment}).

Our goal is to find the optimal values of $\theta_i$ and $\phi_i$ as a function of time and measurement outcomes $\Delta V$ that will maximize entanglement. We choose the fidelity \cite{Jozsa:1994} of $\rho$ with respect to the pure state $\rho_{t0} = \ket{t0}\bra{t0}$, 
i.e., 
${\cal{F}}(\sigma, \rho_{t0}) \equiv \text{Tr}[ \sqrt{ \sqrt{\rho_{t0}} \rho  \sqrt{\rho_{t0}} } ]^2 =  \rho_{t0t0}$,
as a proxy for entanglement rather than the concurrence~\cite{Wootters1998} with this state, since the former yields a figure of merit that is linear in the state $\rho$. We further simplify the setup by enforcing identical local feedback unitaries satisfying the properties $\theta_1=\theta_2$ and $\phi_1=\phi_2=\pi/2$. 
The triplet subspace is closed under local unitary rotations satisfying these constraints. We assume
that the initial state is in the triplet subspace, so that we do not require more the general rotations that connect the singlet and triplet subspaces. This is consistent with initializing in the equal superposition state $|\psi_0\rangle$. 

In order to find the optimal feedback rotation after a measurement, regardless of whether this is in a discrete or infinitesimal time increment, we parameterize the density matrix as follows:
\begin{align*}
&\rho = 
 \begin{bmatrix}
\rho_{t-t-}		& \rho_{t-t0} 		& \rho_{t-t+} 		& \rho_{t-s} \\
\rho_{t-t0}^*	& \rho_{t0t0} 		& \rho_{t0t+} 		& \rho_{t0s} \\
\rho_{t-t+}^*	& \rho_{t0t+}^* 	& \rho_{t+t+} 		& \rho_{t+s}\\
\rho_{t-s}^*	& \rho_{t0s}^*		& \rho_{t+s}^*		& \rho_{ss} \end{bmatrix}
\end{align*}
where $\ket{t-} \equiv \ket{00}$, $\ket{t+} = \ket{11}$ and $\ket{s} \equiv \frac{1}{\sqrt{2}}(\ket{01}-\ket{10})$. Since we are restricting the feedback operations to $\sigma_y$ rotations ($\phi=\pi/2$) and our measurement is represented by a real matrix, we may further restrict $\rho$ to be a real matrix. The fidelity with respect to $\ket{t0}$ after applying identical $\sigma_y$ rotations on both qubits is given by
\begin{align}
{\cal{F}}=  \langle t0|\rho_1&| t0\rangle =  \rho_{t0t0}+\frac{1}{4}(\sqrt{8} \sin 2 \theta (\rho_{t-t0}-\rho_{t0t+})+ \nonumber \\
& (1-\cos 2 \theta ) (1-3 \rho_{t0t0}-2 \rho_{t-t+}-\rho_{ss})),
\label{eq:Fid}
\end{align}
where $\rho_1 = U_F[\theta, \theta, \frac{\pi}{2}, \frac{\pi}{2}]~\rho~ U_F\dg[\theta, \theta, \frac{\pi}{2}, \frac{\pi}{2}]$.
The optimal rotation angle $\theta$ is then found by maximizing $\langle t0|\rho_1|t0\rangle$ over $\theta$ and is given by,
\begin{equation}
\theta_{\text{opt}}[\rho] = \frac{1}{2} \arctan[\sqrt{8}(\rho_{t-t0}-\rho_{t0t+}), 3 \rho_{t0t0}+\rho_{ss}+2\rho_{t-t+} -1],
\label{eq:ThetaOpt}
\end{equation}
where $\arctan[y,x]$ behaves as $\arctan[y/x]$, but with $\theta$ chosen in the correct quadrant, \ie
\begin{equation}
\arctan[y,x] = \left\{ \begin{array}{lr}
					\arctan[y/x] 		& ~ x>0 \\
					\arctan[y/x] + \pi & ~ y \geq 0, x<0 \\
					\arctan[y/x] - \pi	& ~ y < 0, x<0 \\
					(y/|y|) \pi/2		& ~ x=0	
\end{array}
\right. 
\end{equation}

\subsection{\emph{Average-sense} local optimality}
\label{sec:AverageSense}

\erf{eq:ThetaOpt} defines the optimal feedback as a function of the density matrix in a single time step. This relation will define a \emph{locally optimal} protocol (sometimes referred to as a \emph{greedy strategy}
\cite{Dasgupta}), meaning that the controller maximizes the figure of merit at each time step \footnote{The term ``local" refers to time-local here, and not spatially local (as in local unitary rotation).}. However this may not be viable for practical applications. Although the controller does in principle have access to the density matrix at every time step, actually calculating $\rho(t)$, \eg using \erf{eq:RhoMCont} or \erf{eq:RhoM} amounts to dynamical state estimation, which may be too computationally intensive to implement `on-the-fly', i.e., in real time, and hence experimentally infeasible.  Ideally, one would prefer to provide a protocol that does not require storing and manipulating the entire measurement record.

In order to obtain an experimentally feasible protocol, we drop the requirement that a protocol be locally optimal, and instead search for one that is locally optimal in an average sense. Given a Markovian feedback protocol $U_F$, we define the average evolution over a time interval containing measurement and feedback by

\begin{align}
\label{eq:StateUpdateDisc}
& \bar{\rho}(t+\Delta t) = \\
& \int U_F[\bar{\rho}_{\Delta V}(t+\Delta t)] \bar{\rho}_{\Delta V}(t+\Delta t) U_F^\dagger[\bar{\rho}_{\Delta V}(t+\Delta t)] d(\Delta V)
\nonumber
\end{align}
for discrete measurement, and by
\begin{align}
\label{eq:StateUpdateCont}
& \bar{\rho}(t+dt) = \\
&\int U_F[\bar{\rho}_{dV}(t+dt)] \bar{\rho}_{dV}(t+dt) U_F^\dagger[\bar{\rho}_{dV}(t+dt)] d(dV) \nonumber
\end{align}
for continuous measurement. For an ASLO protocol, $U_F[\rho] = U_F[\theta_\text{opt}[\rho], \theta_\text{opt}[\rho], \pi/2, \pi/2]$ and
$\bar{\rho}_{\Delta V(dV)}(t+\Delta t (dt))$ is the average state 
after a measurement initiated at time $t$ as generated by \erf{eq:RhoM} (\erf{eq:RhoMCont}) acting on $\bar{\rho}(t)$. For simplicity of notation, we shall refer to the measurement outcome as $V$, regardless of whether the time interval is discrete or infinitesimal. Note that we are neglecting the time it takes to apply the unitary operations, so that only the measurement contributes to the time duration of the feedback process.

Because of the integration over measurement outcomes, $\bar{\rho}(t)$ is \emph{not} a stochastic quantity, and thus it may be explicitly computed as a function of time for a given feedback protocol 
$U_F[\bar{\rho}_V]$ and initial state $\bar{\rho}(t=0)$. (A controller would only have access to $\bar{\rho}$ if it applied feedback but did not save the measurement record.) 
Because we have set $U_F$ to be a function of $\bar{\rho}_V$, we have implicitly
restricted the feedback operation to depend only on the most recent measurement outcome $V(t)$ and the current time $t$, given the known time evolution of $\bar{\rho}(t)$. Consequently the feedback is Markovian.
The ASLO feedback operation $\bar{\theta}_\text{opt}$ is defined by replacing $\rho$ with $\bar{\rho}_V$ 
in \erf{eq:ThetaOpt}, and can be specified in a lookup table which can be computed beforehand for a given initial state. 
This is a substantial reduction in computational overhead, and as we will see in later sections, this lookup table can in some instances be implemented by a passive device such as a mixer (multiplier), essentially making the feedback autonomous.

It is useful to analyze the symmetries of the framework that we have outlined above.  Suppose we start in an initial state that is symmetric under the transformation $|0\rangle \leftrightarrow |1\rangle$ on both qubits (henceforth referred to simply as a ``01-symmetric'' state). This symmetry implies that $\rho_{t+t+}=\rho_{t-t-}$, $\rho_{t-t0} = \rho_{t0t+}$ and $\langle X \rangle = 0$. Measurement can stochastically break this symmetry by pushing the state toward $|00\rangle$ or $|11\rangle$. However, because the target state $|t0\rangle$ also respects this constraint, feedback will act equally and oppositely for these two cases (\ie $\theta_\text{opt}(V) = -\theta_\text{opt}(-V)$), and thus this symmetry will be restored after integrating over $V$ as in \erf{eq:StateUpdateDisc} or \erf{eq:StateUpdateCont}.
Thus we may assume throughout that $\bar{\rho}$ retains this symmetry so long as the initial state does. Note that unlike both the target and equal superposition state which are both 01-symmetric, the ground state $|11\rangle$ is an example of a state that is not 01-symmetric, and thus we have excluded a natural initial state for the problem.  However in this case it is locally optimal to apply a $\pi/2$ rotation on both qubits, after which we will be in the equal superposition state $|\psi_0\rangle$. 

While feedback strategies designed in this manner may not perform as well as those in which full dynamical state estimation is used at each time step, the added simplicity of average sense local optimality makes the resulting protocols substantially simpler to implement experimentally. 
Note that until Section \ref{sec:Changedt}, none of the strategies we formulate attempt to surpass the performance of a locally optimal protocol. 

\section{Discrete Feedback and the Semiclassical Protocol}
\label{sec:ThresholdFeedback}

We now consider specific feedback protocols based on ASLO introduced in the previous section. In this section we consider the situation where the quantum efficiency of the measurement is small, i.e., $\eta \ll 1$, which is a highly relevant scenario in many experimental settings. In this regime, measurement induced dephasing quickly reduces the off-diagonal elements of the density matrix to zero (see \erf{eq:RhoMCont}), so that the controller only has access to the classical probabilities associated with the three measurement eigenstates. Without knowledge of the coherences, we shall arrive at a semiclassical protocol.  This will not only provide a useful comparison to the more general quantum protocol derived in the following section, but will also be important for developing a hybrid protocol for arbitrary, but not necessary small efficiency $\eta < 1$ (section \ref{sec:Changedt}).

In the limit of very small $\eta$, off-diagonal elements of $\bar{\rho}(t)$ will be approximately $0$. 
To study feedback in this case, we explicitly set the off-diagonal elements to be some small quantities $\rho_{t-t0} = \rho_{t0t+} = \epsilon$ and $\rho_{t-t+} = \epsilon'$ respectively, so that the first argument $(y)$ of the $\text{arctan}$ function in \erf{eq:ThetaOpt}, and thus $\text{arctan}[y/x]$ are approximately zero. If the second argument ($x$) is positive \ie $3\rho_{t0t0, \Delta V}-1-2\epsilon' \approx 3\rho_{t0t0, \Delta V}-1>0$, then $\bar{\theta}_\text{opt}=0$ (assuming for simplicitly of the resulting equations that the singlet subspace is unpopulated). 
Using equation \ref{eq:RhoM}, we see that for an initially 01 symmetric state with fidelity 
$\bar{\rho}_{t0t0}$, the fidelity $\rho_{t0t0, \Delta V}$ after measurement is given by
\begin{align}
\label{eq:IntermediateFid}
&\bar{\rho}_{t0t0, \Delta V} = \\ \nonumber
& \frac{\bar{\rho}_{t0t0}}{\bar{\rho}_{t0t0}+\frac{1-\bar{\rho}_{t0t0}}{2}(e^{-4 \eta k \Delta t(1+2 \Delta V)})(1+e^{16 \eta k \Delta t \Delta V})}
\end{align}

Because \erf{eq:IntermediateFid} decreases monotonically away from $\Delta V=0$, the above inequality yields a threshold behavior for the feedback strategy, in which the preferred operation is to do nothing unless $|\Delta V|$ exceeds some critical value $V_{T,\text{opt}}$. In this case $(x<0)$, then $\bar{\theta}_\text{opt}=\pm \pi/2$ where the sign is chosen to match the sign of $y$. Using \erf{eq:MOp_eta1} to calculate $y$, it is not difficult to show that the sign of $y$ is the same as that of $\epsilon \Delta V$\footnote{If $\epsilon$ is identically zero, then one may take it to be an infinitesimal positive or negative value when calculating $\bar{\theta}_\text{opt}$.}. 
The optimal threshold voltage is given as
\begin{equation}
V_{T,\text{opt}} = \frac{1}{8 \eta k \Delta t} \text{arccosh} \Bigg[ \frac{2 \bar{\rho}_{t0t0}}{1-\bar{\rho}_{t0t0}} \exp(4 \eta k \Delta t) \Bigg],
\label{eq:Threshold}
\end{equation}
which defines the \emph{semiclassical protocol}. Note that $V_{T,\text{opt}}=1/2$ in the projective measurement limit, $k \Delta t \gg 1$.
\erf{eq:Threshold} has a simple interpretation. If the state is already entangled with high probability, the controller does nothing. If the probability of being in either $\ket{t-}$ or $\ket{t+}$ is above a certain threshold, one applies a $\pm\pi/2$ pulse to both qubits, which essentially resets the state to the product state $\frac{1}{2}(|0\rangle + |1\rangle)\otimes((|0\rangle + |1\rangle) = \ket{\psi_0}$ and gives the joint measurement another chance to collapse to the entangled $\ket{t0}$ state. 

This strategy is classical in the sense that the optimal feedback could just as easily be calculated by using the classical Bayes rule to combine prior state knowledge with the information gathered from the previous measurement to determine whether it is beneficial to apply feedback or do nothing~\cite{Korotkov2011}.
Note that the larger $\rho_{t0t0}$
 is, the larger $V_{T,\text{opt}}$ and hence the wider the range of voltage in which no operation is perfomed, $[-V_{T,\text{opt}}, V_{T,\text{opt}}]$. 
Qualitatively, this is because as the fidelity with $\ket{t0}$ increases, it becomes more likely that the qubits are in the entangled state, and so new information needs to suggest with higher probability that the system is in $\ket{t-}$ or $\ket{t+}$ before applying a rotation is beneficial, consistent with the Bayesian interpretation.

If one instead uses a fixed threshold value for feedback, one can analytically solve for the steady state fidelity under this protocol:

\begin{align}
\label{eq:FidSteadyState}
&{\cal{F}}_{ss}= \langle t0|\bar{\rho} (\infty) | t0\rangle =   \\
 &1-\frac{4 \text{erfc}[\frac{V_T}{\sigma}]}{\text{erfc}[\frac{(V_T+1)}{\sigma}]+4 \text{erfc}[\frac{V_T}{\sigma}]+\text{erfc}[\frac{(V_T-1)}{\sigma}]},\nonumber
\end{align}
where $\sigma \equiv 1/\sqrt{4 k \eta \Delta t}$ is the standard deviation of the measurement operator, and erfc is the complement error function, $1-\text{erf}(x)$. We can see by taking limits of this expression that one can only reach unit steady-state fidelity in the limit that $V_T \gg \sigma$, or in other words, the threshold voltage is much greater than the variance of the Gaussian weak measurement effect, \erf{eq:MOp}.

\begin{figure}[h]
    \centering
        \includegraphics[width=0.45\textwidth]{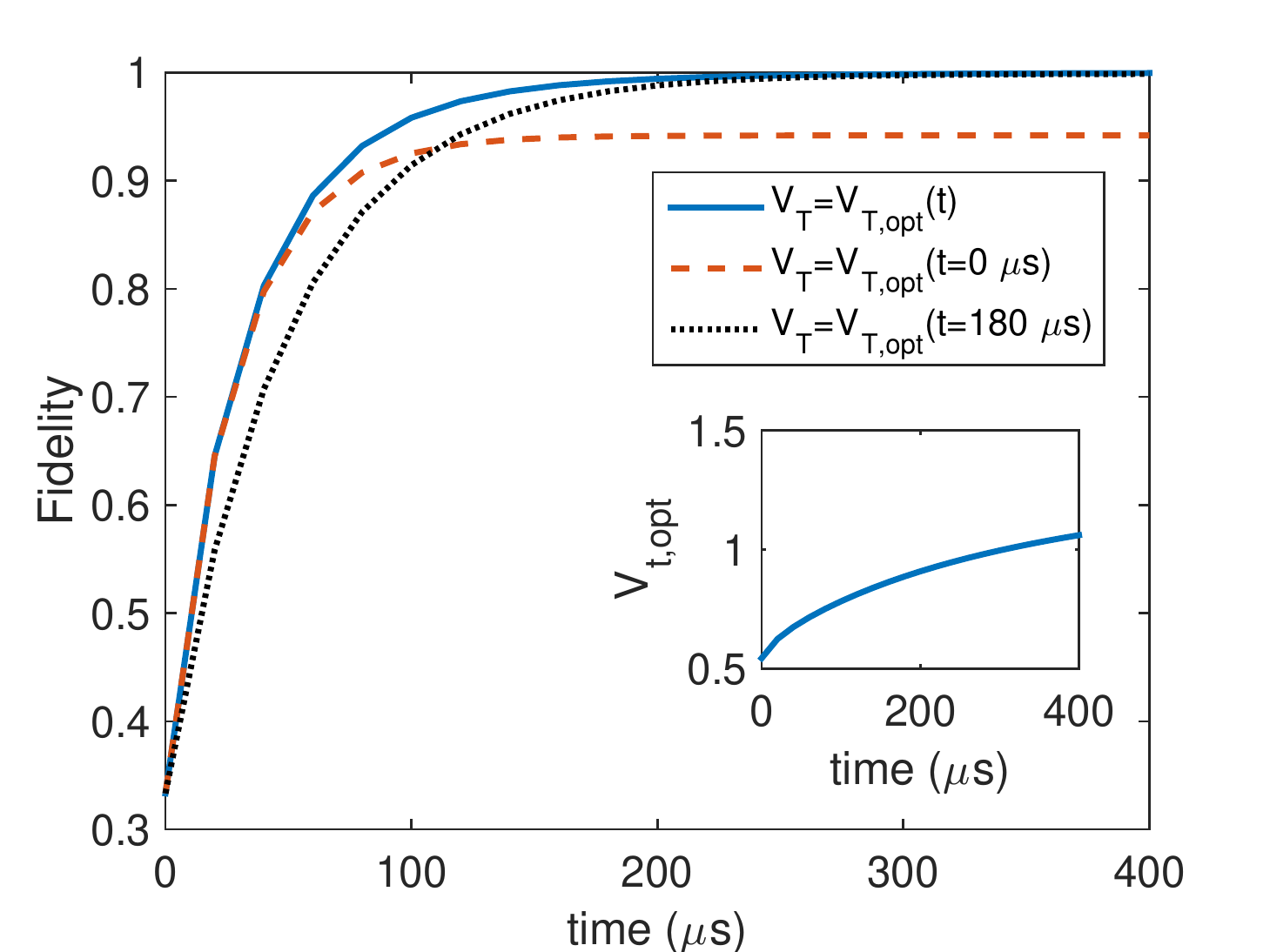}
    \caption{Discrete-time feedback simulations, showing fidelity with the $\ket{t0}$ state as a function of time, starting from the maximally mixed state in the triplet subspace. For these simulations, $\eta=0.1$, $k=1 ~2\pi\text{MHz}$ and $\Delta t = 20 \mu s$. 
The inset shows the optimal threshold voltage as a function of time. Note that for a smaller threshold, fidelity increases quickly at first but then saturates to a value significantly less than one, while for a larger constant threshold, fidelity increases slowly at first but then surpasses the former and approaches unity, though it does not asymptote to $1$ (see \erf{eq:FidSteadyState}). The locally optimal strategy, which increases the threshold as a function of time matches or surpasses both fixed-threshold strategies at all times.} 
   \label{fig:ThresholdFeedback}
\end{figure}

The discussion above defines a viable locally optimal discrete feedback protocol using a half-parity measurement. A similar 
procedure using a full parity measurement  has been used to deterministically entangle qubits located in the same cavity \cite{Riste:2013um}, without analysis or proof of any optimality properties. 

\section{The Continuous-Time Case and the Quantum Protocol}
\label{sec:ContinuousFeedback}

If one attempts to derive a continuous-time protocol using the above result, the increase in fidelity becomes arbitrarily slow in the small $\Delta t$ limit. This is problematic for implementations.  The underlying issue is apparent from examination of the $\Delta t \rightarrow 0$ limit of the threshold voltage in \erf{eq:Threshold} is taken. $V_\text{T,opt}$ diverges as $1/\Delta t$, but the standard deviation of the measurement outcomes diverges more slowly, as $1/\sqrt{\Delta t}$. (We note that the fact that these quantities diverge is an artifact of our normalization convention for $\Delta V$. No physical observable diverges in this limit.) Thus the probability that this feedback strategy will result in performing any operation on the state vanishes. In order to derive a viable continuous-time protocol, we must therefore include coherences and take the full density matrix into account instead of just the diagonal elements.

\begin{figure}[h]
    \centering
        \includegraphics[width=0.45\textwidth]{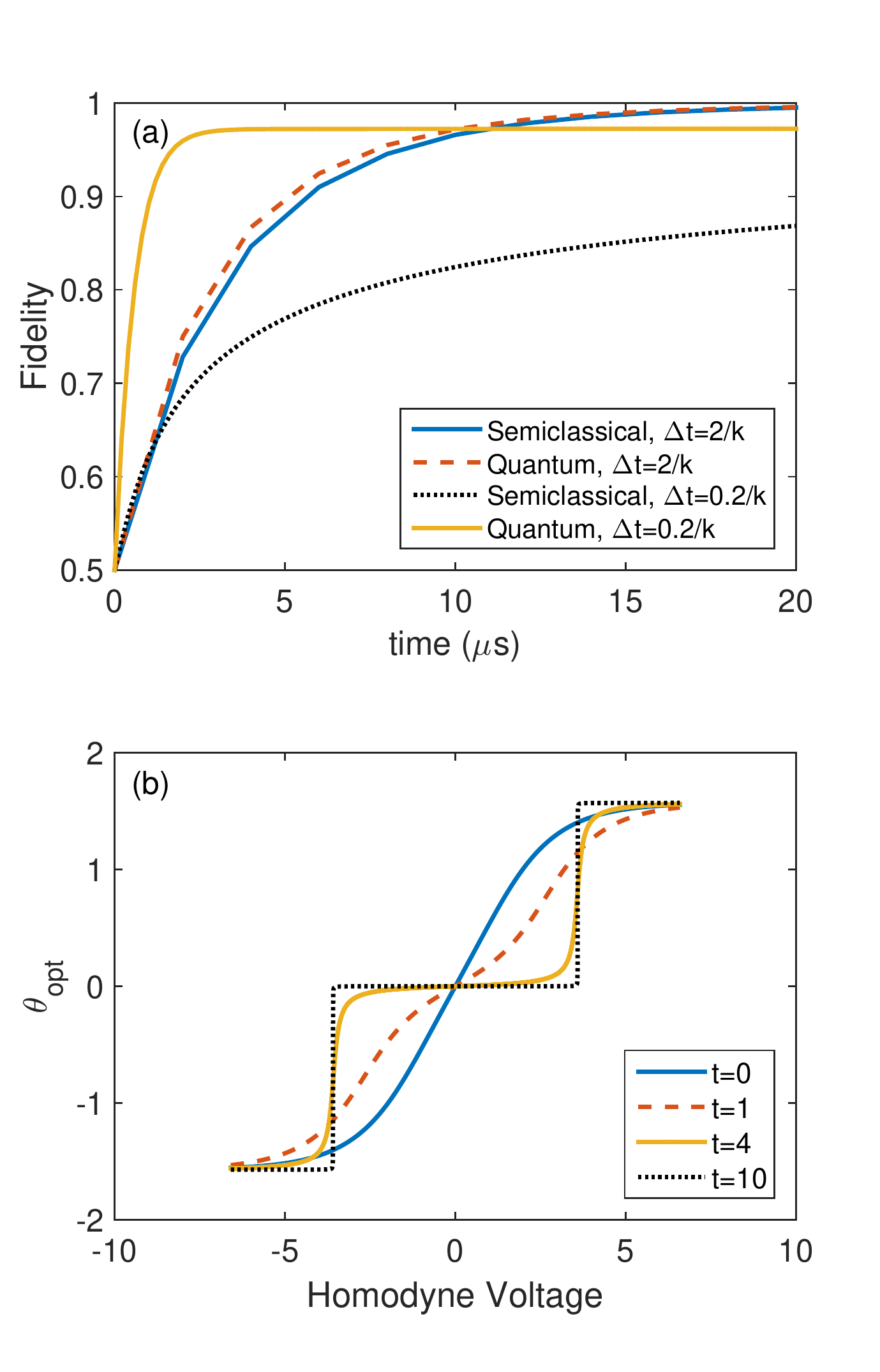}
    \caption{ (a) Discrete feedback simulations, showing fidelity with the $\ket{t0}$ state as a function time under the semiclassical and quantum feedback (\erf{eq:ThetaOpt}) strategies, starting in the 
    equal superposition state $\ket{\psi_0}$ and assuming a unit efficiency measurement and $k=1~2\pi\text{MHz}$. The performance also changes according to the discrete time step, $\Delta t$: we show two representative cases here. (b) The feedback $\theta_\text{opt}(V)$ applied by the quantum protocol shown in panel (a) for the case $\Delta t = 0.2/k$ at four distinct time steps. At the latest time step, $\theta_\text{opt}$ resembles the semiclassical protocol because the off-diagonal elements of the density matrix have decayed to zero. } 
   \label{fig:FeedbackGen}
\end{figure}

We approach this by recognizing that \erf{eq:ThetaOpt} itself defines an ASLO protocol and can therefore be used to derive both discrete and continuous-time feedback strategies directly, without setting the off-diagonal terms of $\rho$ to zero. Fig. \ref{fig:FeedbackGen} shows the performance of this discrete time strategy in the case of perfect efficiency $\eta = 1$ for various choices of $\Delta t$, and compares with the semiclassical protocol derived in the last section.
Not surprisingly, Fig. \ref{fig:FeedbackGen} shows that the quantum protocol has strictly better performance over the semiclassical protocol. 
 However, the performance gap between the two protocols closes as $\Delta t$ increases. The reason for this is that both the measurement-induced dephasing and the time interval between feedback operations increases as $\Delta t$ increases, so that the density matrix for which the feedback is calculated becomes closer and closer to a diagonal form.
One can see this behavior explicitly in Fig. \ref{fig:FeedbackGen} (b), in which we plot the applied feedback as a function of the measurement outcome. At late times, it resembles the semiclassical protocol. We will revisit this point in section \ref{sec:Changedt}.

The above example still constitutes discrete feedback, in which the measurement and feedback unitary rotations act sequentially. To derive a continuous protocol, we take the measurement strength to be small and assume that infinite strength rotations are available.  
We show below that this unphysical  assumption will be unnecessary, but it is convenient for the initial derivation. 

Given that $\bar{\rho}$ is 01-symmetric for any evolution starting from an initially 01-symmetric state, $dV \propto dW$ because $\langle X \rangle=0$. Furthermore, inspection of \erf{eq:RhoMCont} shows
that the quantity $\rho_{t-t0,dV} - \rho_{t0t+,dV}$ appearing in the first argument of the $\text{arctan}$ function in \erf{eq:ThetaOpt} is infinitesimal and proportional to $dW$. We can then substitute the continuous-time measurement update in \erf{eq:RhoMCont} into \erf{eq:ThetaOpt}, and expand in a Taylor series with respect to $\bar{\rho}_{t-t0,dV} - \bar{\rho}_{t0t+,dV} \propto dW$. Assuming that the second argument in the $\text{arctan}$ function is greater than zero\footnote{If we start in the separable state $(|0\rangle + |1\rangle)\otimes(|0\rangle + |1\rangle)$ as is most practical, this constraint will hold true for all later times. If we start in a state that does not satisfy this property, it is locally optimal to first apply a $\pi/2$ rotation on both qubits, after which the second argument will be greater than zero.}, this yields a proportional feedback strategy in which the feedback rotation is proportional to the measurement outcome $dW$ via
\begin{equation}
\bar{\theta}_\text{opt} = \frac{4 \sqrt{k \eta} \bar{\rho}_{t-t0}}{3\bar{\rho}_{t0t0}+\bar{\rho}_{ss}+2\bar{\rho}_{t-t+}-1} dW,
\label{eq:POptTheta}
\end{equation}
or equivalently, 
\begin{equation}
\bar{P}_{\text{opt}}(t) \equiv \frac{\bar{\theta}_{\rm opt}}{dV} = \frac{8 \sqrt{2} k \eta \bar{\rho}_{t-t0}}{3\bar{\rho}_{t0t0}+\bar{\rho}_{ss}+2\bar{\rho}_{t-t+}-1}
\label{eq:Popt}
\end{equation}

This shows that the ASLO feedback is in fact \emph{direct feedback} in the continuous-time limit, where the feedback rotation angle is directly proportional to the measurement value. This is the type of feedback is
modeled by a Wiseman-Milburn feedback master equation \cite{Wis-1994, Wiseman:2009vw} 
given in this instance by
\begin{align}
d\rho = \mathcal{D}[M]\bar{\rho} dt + \mathcal{H}[M]\bar{\rho} \sqrt{\eta} dW - i[H_F, \bar{\rho}] \frac{dW}{\sqrt{8 \eta k}} \nonumber \\
 - i[H_F, \{M,\bar{\rho }\}] \frac{dt}{\sqrt{8k}} + \mathcal{D}[H_F] \bar{\rho} \frac{dt}{8 k \eta},
\label{eq:WMOpt}
\end{align}
with
\begin{equation}
H_F = \frac{\bar{P}_\text{opt}(t)}{2} (\sigma_{y1} + \sigma_{y2}).
\end{equation}
$d\bar{\rho}$ may be calculated as before using the averaged version of \erf{eq:WMOpt} \ie integrating out $dW$.
Substituting \erf{eq:Popt} into \erf{eq:WMOpt} then yields the equations of motion for the state under the \emph{ASLO quantum, continuous-time protocol}.
While this equation is difficult to solve in the general case, for $\eta=1$ it admits an analytic solution when the initial state is pure and satisfies the usual symmetry property. 
To find this solution, we take the trial solution to be pure and 01-symmetric, so that $\bar{\rho} = \ket{\psi}\bra{\psi}$ with
\begin{align}
|\psi(t)\rangle =
\left[\begin{matrix}
\sqrt{\frac{1-\bar{\rho}_{t0t0}(t)}{2}}, & \sqrt{\bar{\rho}_{t0t0}(t)}, & \sqrt{\frac{1-\bar{\rho}_{t0t0}(t)}{2}}, & 0
\end{matrix}\right]^{\mathsf{T}}.
\label{eq:PurAnsatz}
\end{align}
Substituting this form into \erf{eq:WMOpt} yields the following differential equation for $\bar{\rho}_{t0t0}$
\begin{equation}
\frac{d\bar{\rho}_{t0t0}}{dt} = 2k(1-\bar{\rho}_{t0t0}) \implies \bar{\rho}_{t0t0}(t) = 1-(1-\bar{\rho}_{t0t0}(0))e^{-2 k t}
\label{eq:AnsatzSol}
\end{equation}

Crucially, the terms in \erf{eq:WMOpt} that depend on $dW$ cancel upon substitution. This cancellation ensures that the actual state equals the average state at all times \ie $\bar{\rho}(t) = \rho(t)$, so that the average state evolution is pure under this feedback protocol. This implies the non-averaged evolution is also deterministic,
and thus
dynamical state estimation is not necessary to implement the locally optimal strategy in this special case. 

Having a solution for the evolution of the full density matrix in the case $\eta=1$ also yields an analytic solution for the optimal proportionality coefficient between feedback and measurement, ${P}_{\text{opt}}$, as a function of time: 
\begin{equation}
P_\text{opt} (t) = \frac{4k(1-\rho_{t0t0}(0))}{\sqrt{(1-\rho_{t0t0}(0))(\rho_{t0t0}(0)-1+e^{2k t})}}
\label{eq:OptP}
\end{equation}

This equation displays a marked similarity with the optimal feedback for single qubit purification \cite{Jacobs:2003hc}, with the functional form differing only by two minus signs in distinct locations. Like this quantum protocol for pure states, the single qubit state evolution under optimal feedback is also deterministic, and has been shown to be globally optimal for $\eta=1$ \cite{Wiseman:2008bc, Jacobs:2003hc, Li:2013vd}. These parallels lead us to speculate that the protocol given by \erf{eq:OptP} may be globally optimal as well. In the case of non-unit measurement efficiency, $\bar{\rho}_{t0t0}$ appears to exactly follow an exponential which asymptotes to a value less than $1$. However, we have not been able to find an analytic solution. 

\begin{figure}[h]
    \centering
        \includegraphics[width=0.45\textwidth]{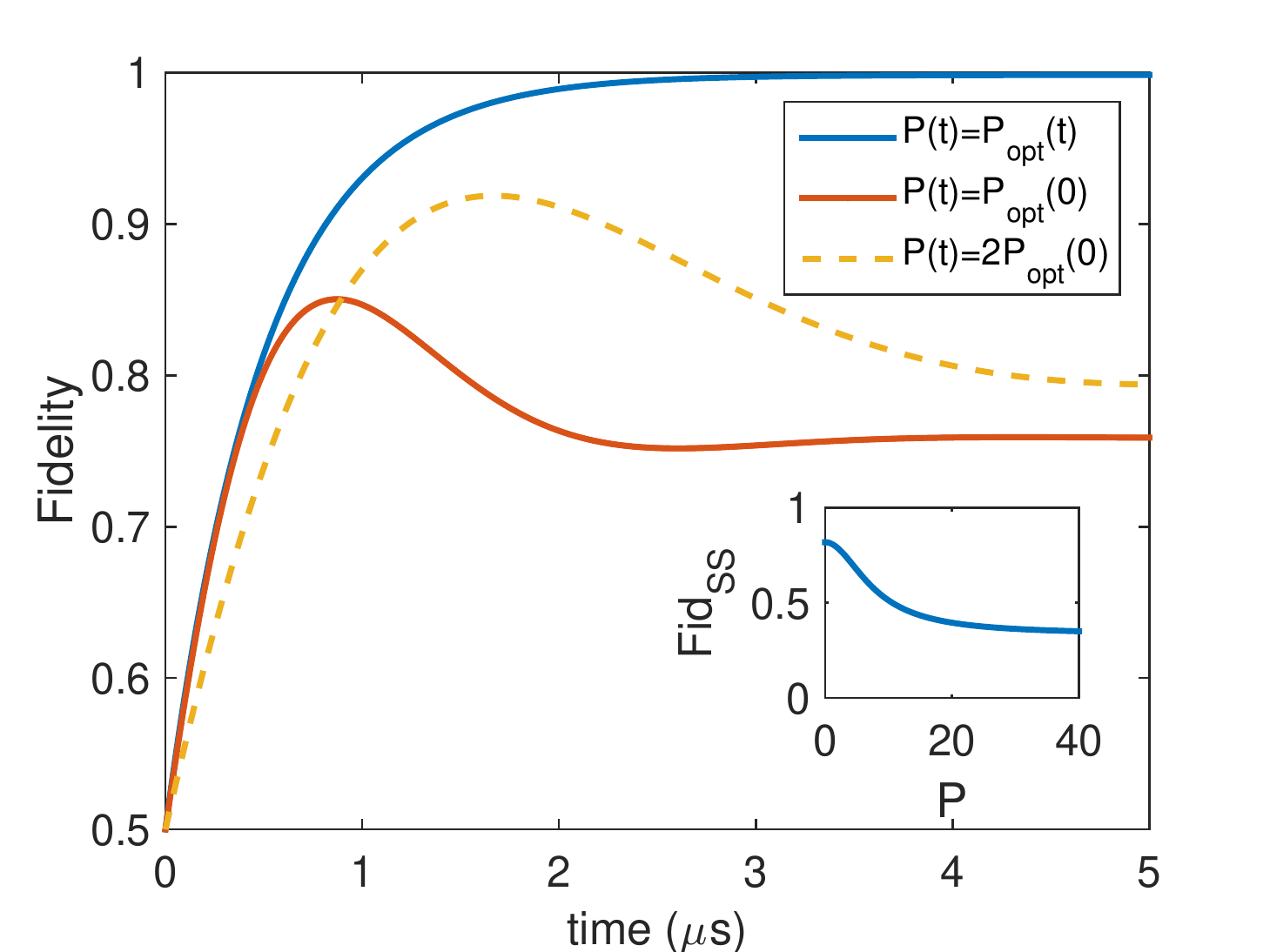}
    \caption{Continuous feedback simulations, showing fidelity with the $\ket{t0}$ state, ${\cal{F}}$, as a function of time under the continuous-time ASLO protocol given in \erf{eq:Popt}. The initial state is the separable state $\ket{\psi_0}$.
    The simulation parameters are $k=1 ~2\pi\text{MHz}$, $\eta=1$, with time step $dt \ll 1/k$. Fidelity as a function of time for several values of constant proportionality coefficient $P$ are also shown for comparison. Inset shows the steady state fidelity ${\cal{F}}_{ss,f}$ achieved for constant $P$, according to \erf{eq:FidSteadyStateP}. Since we have taken the small $dt$ limit, the threshold strategy would not change from its fidelity at $t=0$. Note that if we were to plot the locally optimal strategy in this continuous time situation, which would nomically require dynamical state estimation, it would coincide exactly with the curve for $P_\text{opt}(t)$.}
   \label{fig:POpt}
\end{figure}

Fig. \ref{fig:POpt} shows the fidelity as a function of time using the quantum protocol for $\eta = 1$, comparing the performance of the ASLO protocol with time dependent $P_\text{opt} (t)$ (blue line) to that obtained by using a constant multiple of the zero time value $P_\text{opt} (0)$ (red and yellow-dotted lines). The superiority of the ASLO protocol is evident.

 It is useful to study the asymptotic behavior of these protocols. The numerical simulations show that with the ASLO strategy, the fidelity asymptotes quickly to one, while using a constant multiple of $P_\text{opt} (0)$ does not reach unit fidelity. This raises the question of whether a specific fixed value of $P$ could also yield unit fidelity. By fixing $d\rho=0$ in \erf{eq:WMOpt}, we obtain a system of linear equations which can be solved to yield the steady state fidelity for fixed $P$:
\begin{equation}
{\cal{F}}_{ss,f}= \frac{P^2+16k^2 \eta (1+8\eta)}{3P^2+16k^2 \eta(3+8\eta)}.
\label{eq:FidSteadyStateP}
\end{equation}
This result is valid for $P\neq0$ (when $P=0$, any state that commutes with the measurement is a steady state). For large $P$, the steady state fidelity tends to $1/3$, while in the limit $P \rightarrow 0$, the steady state fidelity is given by $(1+8\eta)/(3+8\eta)$. Moreover, since the denominator in \erf{eq:FidSteadyStateP} is greater than the numerator for all values of parameters, ${\cal{F}}_{ss,f} <1$. This analysis proves that it is not possible to obtain unit fidelity with a constant value of $P$, and therefore a time-varying direct feedback protocol is necessary in order to reach unit fidelity at long times. 

This quantum protocol has several appealing features for experimental realization. Firstly, proportional feedback is realizable simply by using a mixer or analog multiplier \cite{Vijay:2012ua}, both of which have very low latency.
Furthermore, unlike the semiclassical protocol, only infinitesimal rotations are called for in any given time step, which reduces the resources necessary for implementation.

\section{Inefficient measurement and hybrid protocols}
\label{sec:Changedt}

\begin{figure}[h!]
    \centering
        \includegraphics[width=0.45\textwidth]{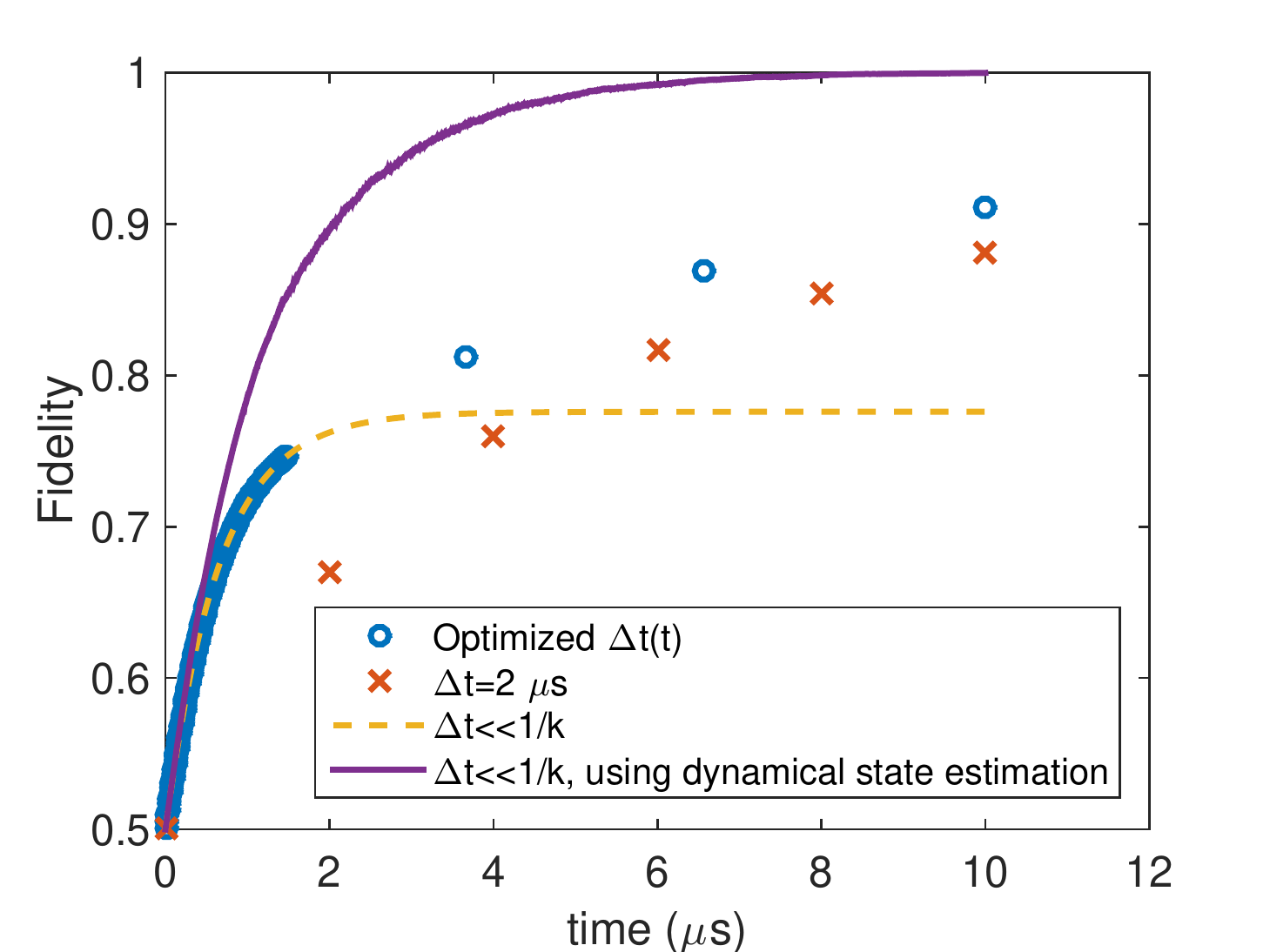}
    \caption{Fidelity ${\cal{F}}$ starting from the $\ket{\psi_0}$ state as a function of time under the optimal variable time step protocol (blue circles), compared to fidelity obtained with constant time step protocols (discrete time semiclassical protocol (red crosses), continuous time quantum protocol (yellow dotted line)).  The measurement is on continuously at a constant rate and feedback is applied instantaneously at each point in the graph. Parameters $k=1 ~2\pi\text{MHz}$ and $\eta = 0.4$ are chosen to illustrate the benefit of this optimized strategy for non-unit but not too small efficiency. The dense cluster of points before $t=2\mu s$ shows that the gradient search resulted in almost all of the times for application of feedback to be located in the early stages of the evolution. Also shown is the fidelity obtained from a full non-Markovian strategy using dynamical state estimation and locally optimal feedback (see text).
    }
   \label{fig:ChangeTau}
\end{figure}

For unit efficiency, the fidelity quickly reaches $1$ when continuous-time feedback is used. But for $\eta<1$, the fidelity asymptotes to a value less than $1$ (see Fig. \ref{fig:ChangeTau}). Qualitatively, this happens because the 
off-diagonal elements of $\rho$
that drive feedback in the continuous-time case (see \erf{eq:Popt}) decay faster relative to the feedback terms which increase the fidelity. In contrast, if we implement discrete-time feedback for the same value of $\eta<1$, the fidelity increases more slowly at first, but eventually surpasses the asymptotic fidelity of the continuous-time strategy. Moreover, we know from Sec. \ref{sec:ThresholdFeedback} that the semiclassical strategy is unaffected by decay of the off-diagonal terms, and this strategy does reach a fidelity of $1$.
However this strategy is only viable when the measurement time is finite. 
These facts suggest that we consider a \emph{hybrid protocol} that transitions between continuous-time and discrete-time feedback; \ie has a variable measurement duration.

To determine how to make this transition, we perform a numerical optimization of the measurement durations as follows. We divide the system evolution into 200 discrete steps each of duration $\Delta t_i$ subject to the constraint $\sum_i \Delta t_i = T_\text{final}$. At each time step, discrete-time measurement, according to \erf{eq:MOp}, followed by feedback according to the general optimal function \erf{eq:ThetaOpt} is applied. Then an optimization over all time intervals $\Delta t_i$ is performed (using the gradient descent algorithm) to minimize the cost function $1-\langle t0|\bar{\rho}(T_\text{final})|t0\rangle$. 

The results of this optimization are shown in Fig. \ref{fig:ChangeTau}, where we plot the resulting fidelity as a function of time, and compare to evolutions in which all $\Delta t_i$ are fixed to a constant
finite or an infinitesimal value. The optimization consistently finds a minimum in which the majority of the $\Delta t_i$ intervals are small and approximately equal, while a few at the end are large. In other words, the optimal solution found by gradient descent shows a sharp transition between continuous-time feedback at short times to discrete-time feedback at long times.
After the switching time, the off-diagonal elements are observed to be small, and thus the applied feedback closely resembles that of the semiclassical protocol.

Since the measurement rate $k$ is held constant in the above optimizations, an alternative way to view the change in optimal measurement time step is to view the finite duration measurements as a series of infinitesimal measurements. This perspective lends itself to the interpretation that when the hybrid protocol uses discrete feedback, it abstains from applying feedback at one time, only to apply a stronger feedback at a later time, and thus is not locally optimal.  While we cannot prove any optimality properties of this hybrid protocol, its superior numerical performance indicates that it is significantly closer to global optimality than both of the fixed $\Delta t$ protocols derived in previous sections.

Fig. \ref{fig:ChangeTau} also shows that there exists a non-Markovian strategy which outperforms the hybrid protocol described above. Simulating continuous feedback and averaging over 300 trajectories, we plot the fidelity as a function of time assuming that the controller performs dynamical state estimation at each time step
and then applies the optimal feedback unitary, \erf{eq:ThetaOpt}. Because simulation of trajectories tracks the actual state as opposed to the average state, this protocol is locally optimal, as opposed to ASLO. This strategy is found to surpass the hybrid protocol at both early and late times, showing that for $\eta < 1$, dynamical state estimation can yield a better protocol than one based only on the average state.
 However, as discussed earlier (see section \ref{sec:AverageSense}), such a non-Markovian protocol is generally more difficult to implement in practice.

\section{Experimental Realization}
\label{sec:Experiment}

In any experimental realization of the proposed feedback schemes to achieve entanglement between remote qubits, numerous imperfections will complicate the dynamics studied above. Firstly, finite coherence of the qubits will limit the fidelity. Second, the above results apply only to truly Markovian dynamics, in which the controller acts instantaneously with an action that is based solely on the most recent measurement outcome. In practice, the homodyne measurement will have finite bandwidth and the feedback will necessarily act with some finite delay.
In this section we present simulations of the continuous-time quantum feedback protocol specified in \erf{eq:Popt}, in which these imperfections are now incorporated. We use the full stochastic master equation including finite detector bandwidth that is derived in \cite{Li2015},
\begin{align}
\label{eq:IFeedback}
&d\rho = \mathcal{D}[M]\rho dt + \mathcal{H}[M]\rho \sqrt{\eta}dW +\mathcal{L}_\text{q}dt \nonumber \\
& -\frac{i}{\tau} [H_F,\rho] \int_{t_0}^\infty e^{-(s-t_0)/\tau} \Bigg[ 2\langle X \rangle(t-s) ds + \frac{dW(t-s)}{\sqrt{\eta}} \Bigg],
\end{align}
where $\mathcal{L}_\text{q}$ models relaxation $T_1$ and dephasing $T_\phi$ on both qubits
\begin{align}
\label{eq:IFeedback_dephasing}
&\mathcal{L}_\text{q} = \sum_i \Big[\frac{1}{2T_{\phi,i}}\mathcal{D}[\sigma_{z,i}]\rho + \frac{1}{T_{1,i}} \mathcal{D}[\sigma_{i}]\rho\Big].
\end{align}
Notice that the term in square brackets in \erf{eq:IFeedback} is simply $dV$. Simulations are for experimental parameters 
$\eta=0.4$, dephasing times $T_{\phi,1} = 6.9 \mu s/2\pi$ and $T_{\phi,2}=30 \mu s/2\pi$ for the first and second qubits, respectively, relaxation time $T_1=20\mu s/2\pi $~\cite{Roch:2014ey}, feedback delay $t_0 = 100 \text{ns}$, feedback loop bandwidth $1/\tau = 1.6 ~2\pi\text{MHz}$~\cite{Vijay:2012ua} and measurement rate $k=1.3 ~2\pi\text{MHz}$.
The effective dephasing times are obtained from combining an intrinsic qubit dephasing time of $30 \mu s/2\pi$ with a loss of $0.04$ in amplitude units between the cavities due to the circulator, which amounts to an effective $\sigma_z$ measurement on the first qubit by the environment \cite{Roch:2014ey}.

To apply the average state feedback protocols, we assume that the effects of delay and finite bandwidth, the non-Markovian effects, are small, so that $\bar{P}_\text{opt}$ is unchanged after neglecting them. In this limit, we can then first run a simulation to calculate the average evolution $\bar{\rho}(t)$ using the continuous time quantum protocol, \erf{eq:StateUpdateCont}. 
This simulation incorporates decoherence but not the feedback delay or finite bandwidth effects. From this first simulation, we extract $\bar{P}_{\text{opt}}(t)$ and we then apply this feedback coefficient to the stochastic simulations using \erf{eq:IFeedback}, which are able to incorporate non-Markovian effects.
This average performance is obtained here by averaging over $1500$ stochastic trajectories. 
For these particular experimental parameters, dephasing prevents the fidelity from reaching the switching point observed above, which was confirmed by applying the same optimization as that of section \ref{sec:Changedt} but including the above dephasing and decay rates. Therefore in this case, our hybrid strategy dictates that  we apply only continuous feedback. 

\begin{figure}[h]
    \centering
        \includegraphics[width=0.45\textwidth]{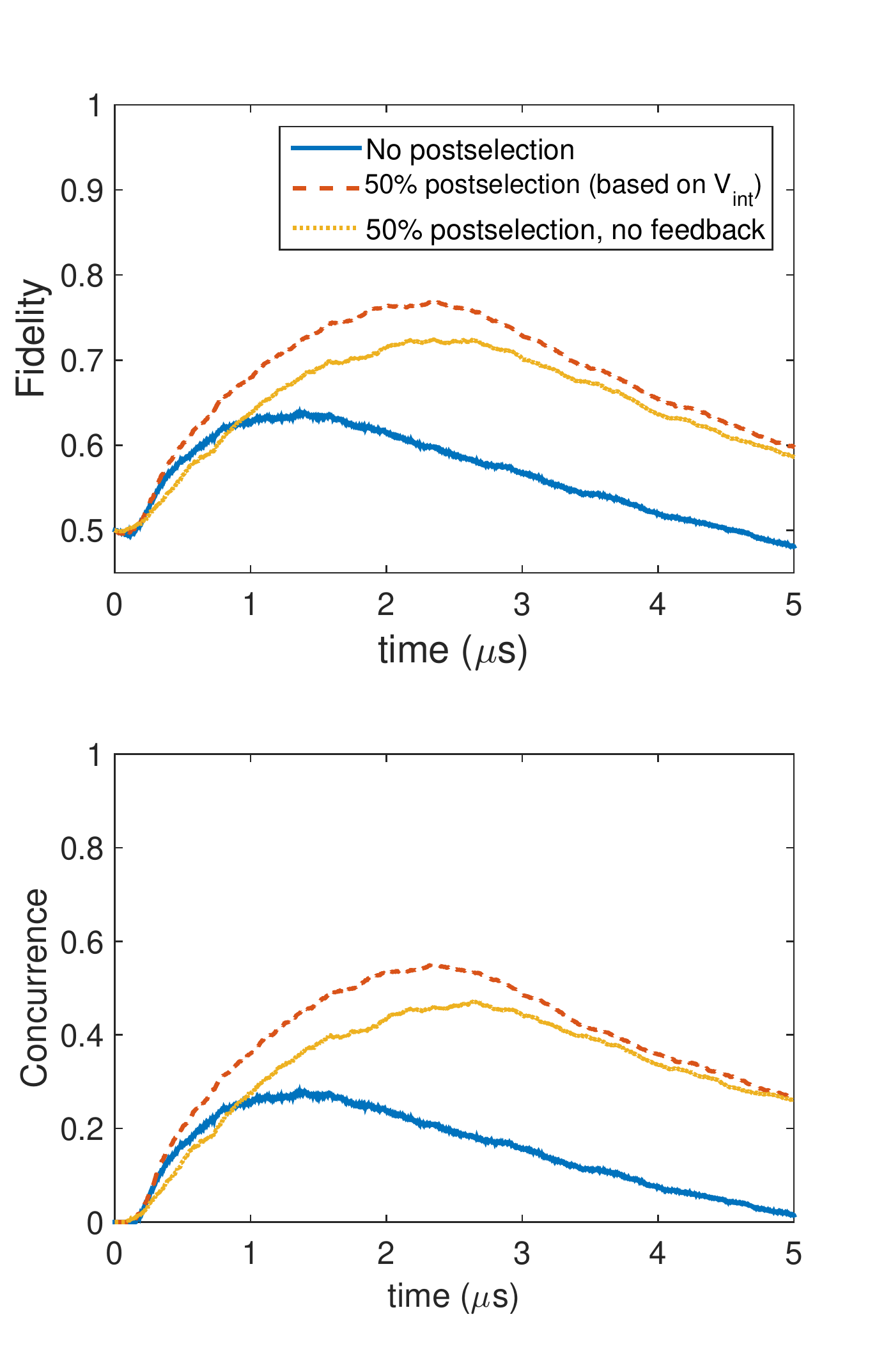} 
    \caption{Continuous feedback simulations, showing fidelity with the $\ket{t0}$ state as a function of time under the continuous-time ASLO protocol prescribed by \erf{eq:Popt}, and incorporating  experimentally realistic parameters and imperfections as described in the text. 
The fidelity reaches a much lower peak value than it would in the ideal case before decaying due to finite dephasing. Despite this, the fidelity substantially surpasses $50\%$, which is the threshold for entanglement. To directly quantify entanglement, we also plot the concurrence \cite{Hil.Woo-1997} as a function of time. Dashed and dotted curves show fidelity using a 50 \% post-selection criteria with and without feedback, respectively.} 
   \label{fig:Stochastics}
\end{figure}

Fig. \ref{fig:Stochastics} shows the time dependence of the fidelity and concurrence resulting from this strategy, i.e., applying the continuous time ASLO ``quantum" feedback protocol in the presence of the seadditional experimental imperfections (solid blue lines).
Figure \ref{fig:Stochastics} shows that existing technology is adequate to implement our proposal and to deterministically entangle two transmon qubits, since both the fidelity and concurrence achieved substantially surpass the corresponding entanglement thresholds to a significant degree. Although the average state feedback protocols derived here are motivated by the goal of achieving deterministic entanglement generation, one can also add post-selection to this protocol even though, unlike the situation in measurement induced entanglement without feedback~\cite{Roch:2014ey}, it is not essential here. 
The red dotted lines in Figure \ref{fig:Stochastics} show the effect on the fidelity and concurrence of using a 50\% post-selection, in which only the trajectories for which the absolute value of the integrated signal from $0$ to $2.7 \mu s$ is less than the median are retained. The value of $2.7 \mu s$ was chose to optimize the peak fidelity. This post-selection method does not require any advanced processing of the signal and is seen to further enhance the fidelity and concurrence. We also plot the fidelity and concurrence using the same post-selection criteria but without feedback, which confirms that feedback results in an improvement.

In the experiment of Ref. \cite{Vijay:2012ua}, the propagation through cables introduced the largest contribution to delay.  Our results show that this delay has a measurable effect on the results of the simulation.
 It is conceivable that the delay could increase as the physical separation between the qubits is increased. It is useful to enquire whether our feedback protocol remains viable in this context. The answer to this question lies in the fact that the measurement strength $k$ effectively sets the timescale for all feedback dynamics, as is evident in \erf{eq:WMOpt}. Making the substitutions $k \rightarrow k/s$, $dt \rightarrow dt s$ and $dW \rightarrow dW/\sqrt{s}$ leads to recovery of the same equation. Thus the effect of feedback delay and finite bandwidth can be removed simply by reducing the measurement strength. This has the effect of slowing down all the dynamics. Of course, one still encounters a limit set by the other absolute time scales inherent to the problem, in particular, the time scales for $T_1$ and $T_2$ processes. Nevertheless, this heuristic scaling argument shows that one can always choose $k$ to strike a balance between delay and dephasing in order to optimize fidelity, and that it is only the effective coherence times that will fundamentally limit the achievable entanglement.

In practice, increasing the distance between the qubits will also increase photon loss between cavities. Since the measurement probe has not yet interacted with the second qubit at this point, loss between cavities introduces additional dephasing on the first qubit. Such dephasing was seen to be significant in the experiment of Ref. \cite{Roch:2014ey}, in which photon loss was dominated by the microwave circulator. Until better circulators are made, this source of loss is also likely to present a significant limitation to the maximum achievable entanglement using this scheme. 

\section{Conclusion}
\label{sec:Conclusion}

In this work we have developed several new protocols for remote entanglement generation using measurement-based feedback that do not require real time quantum state estimation.  We introduced the notion of average sense locally optimal (ASLO) protocols, in which the feedback operations are determined at a specific (local) time by maximizing the fidelity of the average state after a discrete or infinitesimal measurement with respect to the target state.  Using this approach, we derived a quantum feedback scheme with local unitary operations that allows deterministic generation of entanglement when the measurements can be performed with unit efficiency ($\eta = 1$). The time local measurement averaging results in a Markovian feedback that is applicable to both discrete and continuous time implementation.  In the continuous time limit,  the optimal ASLO quantum feedback becomes equivalent to simple proportional feedback, which is easily modelled using a Wiseman-Milburn equation~\cite{Wis-1994, Wiseman:2009vw} and which may be realized in an autonomous fashion in experiments using   a mixer (multiplier) \cite{Vijay:2012ua}. The ASLO strategy was then used to develop a discrete time step, semiclassical protocol, suitable for low measurement efficiencies and large time steps, in which only the classical probabilities for being in the entangled or unentangled states are taken into account.  Analysis of the asymptotic behavior of both quantum and semiclassical protocols led to the development of a hybrid strategy which transitions from the quantum protocol at early times to the semiclassical strategy as the target state is approached at longer times, with an accompanying change in time step at the switching point.  We demonstrated that such a hybrid strategy can be beneficial for the general case of intermediate measurement efficiency, $\eta < 1$, with numerical optimizations and found strong evidence that the hybrid strategy is significantly closer to global optimality than any fixed time increment ASLO protocol.

The ASLO feedback strategies developed here possess interesting relationships to the locally optimal strategies for single qubit purification by measurement and feedback~\cite{Li:2013vd}. In particular, the semiclassical protocol bears some resemblance to the locally optimal strategy for qubit purification in the small $\eta$ limit, while
the quantum protocol at unit measurement efficiency also bears a striking resemblance to the corresponding optimal feedback for qubit purification in Ref.~\onlinecite{Li:2013vd}.  The advantages of combining two different strategies tailored to different measurement efficiencies in a single hybrid strategy were also observed in single qubit purification, although here we have taken the additional step of optimizing the transition time.

We investigated the performance of these ASLO protocols for generation of remote entanglement between superconducting qubits with calculations based on existing superconducting 3D transmon technology. We found that even under the current conditions of relatively low measurement efficiency, the ASLO protocols can deterministically generate amounts of entanglement substantially surpassing the entanglement threshold \cite{Wootters1997}. In contrast, the known methods of heralded entanglement based only on measurements are always probabilistic. We further showed that in the presence of low measurement efficiency, one can also use post-selection to further enhance the fidelity and we therefore developed a simple scheme based on the integrated signal to further enhance the fidelity to the target entangled state.  This analysis highlights an interesting advantage of this remote entanglement generation scheme over others, such as those based on single photon counting and the Hong-Ou-Mandel effect \cite{Hofmann2012, Bernien2013}. 
The ability to use feedback in our setup allows one to enhances the probability of success in a way that is not possible in these other heralded entanglement schemes. While the analysis presented in this work most immediately applies to superconducting qubits in microwave cavities, it could of course be adapted to any system in which one can implement a weak measurement of the half-parity observable defined in section \ref{sec:Measurement}.

This work suggests several future avenues for research that would be interesting from both a theoretical and an applied point of view.
First, one could apply the verification theorems to address the question of whether or not our protocols are globally optimal \cite{Shabani:2008}. In the case of unit efficiency and no decoherence, we speculate that our solution is globally optimal, although we have not attempted to prove this statement. The deterministic evolution of the state under the quantum ASLO protocol indicates promise for this, in part because this feature, i.e., the deterministic evolution, is also observed in the globally optimal protocol for single-qubit purification. Second, while fidelity is a useful figure of merit for generating known states, concurrence may be more suitable for a detailed theoretical study and could yield additional insights. For instance, unlike fidelity, concurrence is invariant under the allowed feedback Hamiltonians, mirroring the behavior of entanglement. It would be useful to see whether this shared symmetry could be exploited.
Finally, while we have ignored dephasing in the development of the ASLO feedback protocols, in applying the protocols to the realistic experimental setting for superconducting transom qubits, we found that dephasing presented the main limitation to performance. This suggests that it would be useful to investigate related protocols that would focus on correcting dephasing errors, and hence assist with remote entanglement stabilization.

\begin{acknowledgements}
We thank Mollie Schwartz and Irfan Siddiqi for many useful discussions. The effort of LM was supported by grants from the National Science Foundation and the Berkeley Fellowship for Graduate Study.
Sandia is a multi-program laboratory managed and operated by Sandia Corporation, a wholly owned subsidiary of Lockheed Martin Corporation, for the United States Department of Energy's National Nuclear Security Administration under contract DE-AC04-94AL8. 
\end{acknowledgements}

\appendix
\section{Derivation of finite time POVM}
\label{app:effect}
As discussed in the main text, the stochastic master equation (SME) associated with a continuous-time QND measurement of the Hermitian observable $X$ with strength $k$, on a system with free Hamiltonian $H$ is:
\beq
d \rho  = -i[H,\rho] dt + 2k\mathcal{D}[X]\rho dt + \sqrt{2k}\mathcal{H}[X]\rho dW(t).
\label{eq:sme}
\eeq
The superoperators $\mathcal{D}$ and $\mathcal{H}$ are defined in the main text. We assume that the measurement is unit efficiency, and since it is QND,  $[H,X]=0$. This equation describes the evolution of the system conditioned on the measured voltage
\beq
dV(t) = \expect{X}(t)dt + \frac{dW(t)}{\sqrt{8k}}.
\eeq

The \emph{linear} stochastic master equation \cite{Goe.Gra-1994, Wis-1996, Jac.Kni-1998} associated with this equation is:
\beq
d \brho  = -i[H,\brho] dt + 2k\mathcal{D}[X]\brho dt + \sqrt{2k}\bar{\mathcal{H}}[X]\brho dW(t),
\label{eq:lsme}
\eeq
with $\bar{\mathcal{H}}[A]\brho \equiv A\brho + \brho A\dg$. While \erf{eq:sme} is nonlinear in $\rho$ and produces a normalized density matrix (\ie $\textrm{tr}(d \rho) = 0$), \erf{eq:lsme} is linear in $\brho$ and produces an unnormalized density matrix. Further, since this equation is linear we can restrict our focus to its action on pure states (since any density matrix is a convex sum of pure state density matrices). This defines the associated linear stochastic Schr\"odinger equation (SSE) \cite{Goe.Gra-1994}:
\beq
d\ket{\bpsi} = \left[ \left(-i H - k X^2\right) dt + \sqrt{2k}dW(t) X\right] \ket{\bpsi}.
\label{eq:lsse}
\eeq
This is again a linear equation in the state, $\ket{\bpsi}$.

Both the linear SME and linear SSE sacrifice the normalization of the resulting state for linearity. To see what this means physically, note that both equations are consistent with $\expect{X}=0$, and hence $dV(t) = \frac{dW(t)}{\sqrt{8k}}$, which means that we are generating conditional dynamics according to statistics associated with some fictitious (time-independent) state with property $\expect{X}=0$. The real state $\ket{\psi}$ may not have this property and this is why the normalization is incorrectly predicted by \erf{eq:lsse}. 

Despite this issue with unnormalized states, the utility of \erf{eq:lsse} is that it is sometimes possible to analytically solve for $\ket{\bpsi}$, see for example Ref. \cite{Jac.Kni-1998}. In fact, the linear SSE in \erf{eq:lsse} is in the easiest class of such equations to solve since all the operators in it commute. We are interested in the solution  to \erf{eq:lsse} over a small finite  
measurement time $\Dt$.  This can be solved explicitly and is given by \cite{Jac.Kni-1998})
\bqa 
\ket{\bpsi_\Dw(t+\Dt)} &=& e^{-iH \Dt} e^{-2k X^2 \Dt + \sqrt{2k}X \Dw} \ket{\bpsi(t)} \nn \\
&=& e^{-iH \Dt} e^{-2k X^2 \Dt + 4kX \Delta V} \ket{\bpsi(t)} \nn \\
&\equiv& \bar \Omega_{\Delta V} \ket{\bpsi(t)},
\label{eq:lsse_soln}
\eqa 
for any initial state $\ket{\bpsi(t)}$ (including normalized states), and where $\Dw = \int_{t}^{t+\Dt} dW(t')$, is a Gaussian random variable. In the second line we have used the relation between the measured voltage and $\Dw$ to write the solution in terms of the measured quantity, $\Delta V$, which has probability distribution
\beq
P_{\Delta V} = \sqrt{\frac{4k}{\pi\Dt}} e^{-\frac{4k \Delta V^2}{\Dt}}.
\eeq
\erf{eq:lsse_soln} generates states that do not quite yield correct predictions since the normalization of the state is incorrect. Consider a normalized initial state $\ket{\psi(t)}$. Then the probability of evolving  
during the measurement time according to the stochastic 
variation $\Delta V$ to time $t+\Dt$ is not actually $\langle \bpsi(t+\Dt) | \bpsi(t+\Dt) \rangle$, but rather
\cite{Goe.Gra-1994, Wis-1996,Jac.Kni-1998}
\beq
P_{\Delta V} \langle \bpsi(t+\Dt) | \bpsi(t+\Dt) \rangle = P_{\Delta V} \bra{\psi(t)} \bar{\Omega}_{\Delta V}\dg \bar{\Omega}_{\Delta V} \ket{\psi(t)}. \nn
\eeq
This expression immediately tells us that the 
measurement operator associated to the true finite time evolution is
\beq
\Omega_{\Delta V} = \sqrt{P_{\Delta V}} \bar{\Omega}_{\Delta V},
\eeq
since the true probability of this stochastic evolution is $\bra{\psi(t)} E_{\Delta V} \ket{\psi(t)}$ for the positive-valued effect $E_{\Delta V} = \Omega_{\Delta V}^{\dagger}\Omega_{\Delta V}$.

Now, solving for this effect for the QND measurement, we get
\bqa
\Omega_{\Delta V} &=& \left(\frac{4k}{\pi\Dt}\right)^{\frac{1}{4}} e^{-iH \Dt} e^{-\frac{2k \Delta V^2}{\Dt} -2k X^2\Dt + 4k X \Delta V} \nn \\
&=& \left(\frac{4k}{\pi\Dt}\right)^{\frac{1}{4}} e^{-iH \Dt} e^{-2k\Dt\left(\frac{\Delta V}{\Dt} - X \right)^2}.
\eqa
$\Omega_{\Delta V}$ is the finite-time measurement operator associated with the $\eta=1$ weak QND measurement used in the main text (with $H=0$), and $E_{\Delta V} = \Omega_{\Delta V}^{\dagger}\Omega_{\Delta V}$ is its corresponding effect.

For completeness we can also show that the effects defined above constitute a POVM since,
\bqa
\int_\infty^\infty d(\Delta V)~ dE_{\Delta V} &=& \int_\infty^\infty d(\Delta V)~ \Omega_{\Delta V}\dg \Omega_{\Delta V} \nn \\
 &=& \int_\infty^\infty  d(\Delta V)~ P_{\Delta V} e^{-4k X^2 \Dt + 8k X \Delta V} \nn \\
&=& \sqrt{\frac{4k}{\pi\Dt}}  e^{-4k X^2 \Dt}\int_\infty^\infty d(\Delta V)~ e^{-\frac{4k \Delta V^2}{\Dt} + 8k X \Delta V} \nn \\
&=& \sqrt{\frac{4k}{\pi\Dt}}  e^{-4k X^2 \Dt} \sqrt{\frac{\pi\Dt}{4k}}  e^{4k X^2 \Dt}  \nn \\
&=& \mathbf{1}.
\eqa 
Here the integral on the third line is a Gaussian integral that can be evaluated by completing the square in the exponent. 

\bibliography{HPF_bib}

\clearpage
\end{document}